Metacognitive Training in Professional Development

Can Improve and Sustain Student Achievement

Jeffrey A. Phillips, M. Catherine McElwain and Katharine W. Clemmer

Loyola Marymount University

Abstract

Secondary school students in the United States continue to underachieve in mathematics and science. Improving teacher quality is a core component of improving student achievement. Here we report on a professional development program, the MAST System, that develops the knowledge and skills for teaching mathematics, including metacognitive knowledge and regulation. In this cognitive apprenticeship program, teachers learn to plan, evaluate and adjust to improve student engagement and achievement. Central is the metacognitive practice of defense of instruction. By practicing this reflective approach, teachers become adaptive experts, able to innovate in the classroom. During the two-year intervention, the MAST System resulted in large increases in the California Standards Test mathematics scores, compared to both the district and the state. In addition, improvement continued for several years after the intervention was completed. This continued improvement in student scores indicated that the teachers and schools changed in a sustainable way.

**Introduction**

The U.S. education system fails to teach mathematics and science adequately to students in K-12 (National Academy of Sciences [NAS], National Academy of Engineering [NAE], Institute of Medicine [IOM], 2007 & 2010; ACT, 2015). The US ranked 36th among 65 economies surveyed on the 2012 Organisation for Economic Co-operation and Development (OECD) Programme for International Student Assessment (PISA) exam, and the scores for US students were statistically lower than the average for all OECD economies (Organisation for Economic Co-operation and Development, 2014). These results are similar to previous years (Hiebert et al., 2003; Mullis, Martin, Foy, & Arora, 2012). While there have been some gains in math proficiency since the 1990s, the rate of improvement has slowed in recent years and reversed in 2014 (http://www.nationsreportcard.gov; DeSilver, 2015).

Most Americans believe that the US education system performs poorly in math and science education. In the 2014 Pew Trust survey only 29% of the general public and 16% of the scientific community (scientists belonging to the American Association for the Advancement of Science (AAAS)) rank US K-12 education as above average in science, technology, engineering and mathematics (STEM) disciplines. In fact, more scientists rank US K-12 STEM education as below average (46%) than above average. This is in spite of the widely held conviction that US scientific achievement is the best in the world (92% of AAAS scientists and 54% of the general population) (Funk, Rainie & Page, 2015).

Poor preparation in mathematics and science affects the number of students entering the STEM disciplines in college and the size and quality of STEM-trained workforce (NAS, NAE & IOM, 2007, 2010; ACT, 2015). Policy-makers in Washington (Mervis, 2009; U.S. Congress Joint Economic Committee [JEC], 2012) and the scientific community (Singer, 2009) worry that

U. S. high school graduates are not only unprepared to enter the STEM disciplines in college or fill STEM jobs in the workplace, they are unprepared to meet the challenges of 21st century life (NAS, NAE & IOM, 2007, 2010; JEC, 2012; Olson & Riordan, 2012).

Teacher quality is the single largest contributor to student achievement in mathematics and science (Goldhaber & Brewer, 1997). The number of high quality teachers available in the STEM disciplines, however, may be insufficient to meet the current and projected needs. (Committee on STEM Education, 2013; JEC, 2012). In his 2011 State of the Union address, President Obama set as a goal 100,000 new STEM teachers in 10 years (The White House, Office of the Press Secretary, 2011).

There are three common approaches to improving the pool of STEM teachers, succinctly summarized by then Secretary of Education Arne Duncan in his interview with Science in April 2009 (Mervis, 2009). One approach is to increase the number of well-qualified STEM graduates who choose K-12 education as a career. This includes programs like Teach for America (http://www.teachforamerica.org) and Math for America (http://www.mathforamerica.org) and programs that provide incentives, such as loan forgiveness, to students so they will enter teaching (see, for example, the U.S. Department of Education, Science, Technology, Engineering and Math: Education for Global Leadership, http://www.ed.gov/stem). A related approach is to make it easier for mid-career STEM professionals to enter education. This can be done by lowering or even eliminating barriers to credentialing for qualified STEM professionals (for review of some programs see Price, 2013).

These two approaches are predicated on two assumptions. First, better preparation in mathematics and science alone will increase the quality of STEM teaching. Although there is a generally positive relationship between the academic preparation that a teacher has in STEM and student achievement in those disciplines, the effect is strongest for secondary math, weaker for secondary science, and interchangeable with academic preparation in education in elementary school (Bolyard & Moyer-Packenham, 2008). The second assumption is that the shortage of qualified teachers in STEM is primarily a pipeline problem. Ingersoll and Perda (2010) suggest that the number of teachers entering STEM teaching is sufficient and that the problem lies in retention and distribution. However, there is little empirical evidence that either of these approaches has the desired effect on student performance (Darling-Hammond, Holtzman, Gatlin, & Heilig 2005; Laczko-Kerr & Berliner, 2002; Hill et al., 2008). Some evidence even suggests that alternative credentialing has a negative impact on student achievement in mathematics and science (Laczko-Kerr & Berliner, 2002; Moyer-Packenham, Bolyard, Kitsantas & Oh, 2008).

      The third approach for improving the pool of STEM teachers is increasing the quality of the teachers already in the profession. In-service professional development (PD) is an attractive model for addressing the need to increase teacher quality because it uses the teacher pool already in place in the schools. This preserves a valuable manpower resource that is already committed to K-12 education. Furthermore, high quality PD is correlated to reducing turnover among STEM teachers (Ingersoll & Perda, 2010; Ingersoll & May, 2012). Unlike the first two approaches, which impact only STEM subject mastery, PD can address not only discipline specific knowledge but also the specialized content knowledge required to teach a STEM discipline like mathematics, the Mathematical Knowledge for Teaching or MKT (Hill et al., 2008). An apprenticeship model focused on in-service teachers in their own classrooms may be the most effective way to address all the kinds of knowledge and skills a STEM teacher requires to increase student achievement.

Accumulating evidence suggests that PD can be an effective tool for increasing the quality of STEM teachers but the effects are not robust or universal (Bolyard & Moyer-Packenham, 2008; Quint, 2011; Yoon, Duncan, Lee, Scarloss, & Shapley, 2007; Sztajn, 2011; Loucks-Horsley, Stiles, Mundry, Love, Hewson, 2010; Akiba & Lang, 2016; Wei, Darling-Hammond, Andree, Richardson, Orphanos, 2009; Darling-Hammond, Wei, Andree, Richardson, & Orphanos, 2009). What constitutes success is not even well-defined (Bolyard & Moyer-Packenham, 2008). Only rarely do studies track the effect on student achievement directly (Akiba & Liang, 2016; Wei et al., 2009). Yoon et al. (2007) reviewed over 1,300 studies on the effect of professional development. Of these, only nine met the What Works Clearinghouse evidence standards (U.S. Department of Education, Institute of Education Sciences, 2013). Of these only five included mathematics and science. In these five studies fifteen effects of PD are reported. Of these, only three, all showing a positive effect, were statistically significant (Yoon et al., 2007).

Guskey and Yoon (2009) identify three important components common to the studies with a positive impact (Yoon et al. 2007). First, successful programs meet for thirty hours or more. Second, the successful programs all have structured and sustained teacher assistance. Third, all the successful programs focus on specific subject content or subject-related pedagogy. It is noteworthy that the activities included in successful programs varied widely; however, the successful programs all included some kind of workshop or summer institute utilizing experts.

In order for in-service PD to impact student achievement long term, it must help teachers improve instructional practice in their own classroom and also institutionalize the culture of teacher learning (Wei et al. 2009), but this kind of sustainability is rare (Quint, 2011). Effective PD must address the needs of different teachers in different teaching situations. Both Farmer,

Gerretson and Marshall (2003), and Koellner and Jacobs (2015) argue that PD systems incorporating metacognitive reflection and higher levels of adaptive learning are more effective than those with a more specified style (see also Anthony, Hunter & Hunter, 2015). This metacognitive reflection allows a teacher to be successful in the ever-changing landscape that teachers encounter. Teachers face new challenges with each new class, each change in content or each new government mandate. In fact, the landscape changes daily as the student needs change. Teachers cannot blindly apply past experience; rather, they must constantly re-evaluate, be flexible and innovative and create new solutions. Successful innovation requires metacognition, or self-regulation (Flavell, 1979; Zimmerman, 1989).

As described by Hatano and Inagaki (1986), teachers who effectively employ metacognition or self-regulation in their teaching practice can be described as adaptive experts, rather than routine experts. Routine, or classic, experts are proficient in the core procedures but lack the skills to innovate and create new solutions. Teachers with adaptive expertise can adjust their classroom practices as needed. Teachers need to be aware of their own, and their students', knowledge and skills (metacognitive knowledge) as well as know how to regulate their own, and their students', learning through intentional actions (metacognitive regulation) (Flavell, 1979). Teachers with these skills are well-equipped to design and implement classroom differentiation, accommodations, and interventions that are developmentally and culturally responsive to the students' needs and learning goals. This adaptive expertise is the "gold standard" in the profession, employing metacognitive strategies, such as research-supported planning, data-driven analysis, and appropriate adjustments (Hammerness, Darling-Hammond & Bransford, 2005). To help teachers be successful in their changing landscape, professional development providers need to develop not merely routine experts but rather adaptive experts.

In this paper we describe an effective STEM PD program, demonstrate that it successfully improved student performance in mathematics, provide evidence that growth continued after the intervention, and suggest that focusing on approaches that develop metacognitive skills, and therefore adaptive experts, was the source of its success. In the following sections, we discuss the various components of this successful PD program. The curriculum is comprised of four modules that were taught over a period of two years. This curriculum was taught via a cycle of seminars, workshops, classroom observations, and debriefing sessions. Central to all of these activities was a metacognitive exercise called Defense of Instruction, in which teachers critically analyze their decisions and reflect on their, and their students', progress. This process was vital to developing the adaptive skills and habits necessary for teachers to become adaptive experts. We demonstrate that the MAST System was effective in raising the scores of students on the California Standards Test. This increase is shown to persist after the formal intervention has concluded, which indicates that the teachers and schools have changed in a sustainable way.

**The MAST System**

The Mathematics and Science Teaching (MAST) System was developed as part of the Center for Math and Science Teaching at Loyola Marymount University (LMU) and is now the foundation for LMU's Teacher Leadership by Design Program (http://www.mathleadershipcorps.org). MAST was developed in 2002, piloted in a single school (2003-2005) and then expanded to multiple schools in 2005-2009. MAST has been used in almost thirty schools. The MAST System was designed using research-based practice for effective professional development including, for example, prolonged duration, subject-related pedagogy, adaptable structure and extensive support, as described above. LMU faculty from both the College of Science and Engineering and School of Education worked together to provide support for the teachers.

The two-year program included on and off-site workshops, discussions, in-class observation, coaching and evaluation. The MAST design included educational theory and discussion of conceptual applications but the focus was always on the teachers' own classroom practices (Borasi & Fonzi, 2002). Teachers used their own data on student engagement and achievement to reflect on their classroom practices. This personalized reflective process developed a continuous cycle of improvement (Learning Forward, 2008). Clinical faculty, who continued to teach in secondary classrooms, served as coaches. These coaches modeled and provided the scaffolding the necessary metacognitive knowledge and regulation as well as offered individual feedback, helping teachers move toward adaptive expertise. By focusing on developing metacognitive skills, the MAST System resulted in improved student performance scores both during and after the two-year intervention.

**MAST Curriculum**

The *MAST Curriculum* was comprised of four Modules taught over a period of two years covering topics drawn from research on how people learn (Bransford, Brown, & Cocking, 1999). This research informed not only the content but also the structure of the entire MAST curriculum, so that the curriculum itself provided a working model of these concepts. On-going evaluation was woven into the metacognitive strategies and evidenced in the data collected in the classroom. While the goal was for teachers to progress through two modules each year, this was not a rigid requirement. Teachers who were not prepared to move to the next module, repeated a module. Readiness to progress was determined by the teacher's score on the MAST Module

Rubric. This created a differentiated or adaptable program responsive to the individual needs of a particular teacher.

The first two modules focused on developing the teacher's skills as a routine expert (see Hatano & Inagaki, 1986). This knowledge acquisition was balanced with developing the metacognitive skills required for planning, evaluating and adjusting those strategies to a specific classroom. The second two modules focused on deepening these metacognitive skills toward developing adaptive experts (see Hatano & Inagaki, 1986).

The first module introduced teaching and learning strategies for engaging students and deepening mathematical understanding. This module included strategies for engaging students before a lecture by connecting student experience to the learning culture of mathematics, using an activity to appropriately connect past and present mathematical content before introducing new concepts, and utilizing multiple assessment techniques to gauge the breadth and depth of knowledge. The teachers assembled these components into effective lesson plans that they implemented in their own classes. In this way, teachers were able to adapt the strategies to best fit their experiences and those of their students. The MAST System staff coached teachers in making informed decisions on implementation based on their individual context. This continual critical reflection began to establish the habits of an adaptive expert. In this module, teachers also began to generate data, collected both formally and informally, to assess the success of these strategies.

The second module focused on establishing a community of teacher learners to help teachers engage productively with their colleagues. Collaboration enabled teachers to learn from the insights, successes and struggles of others and also facilitated goals setting, instructional change, and consideration of other perspectives (Lomos, Hofman, & Bosker, 2011; Vangrieken,

Douchy, Raes, & Kyndt, E., 2015, for reviews). In addition to cultivating a collaborative culture among teachers, in Module 2 teachers were provided with the strategies to involve students in collaboration including the assessment process. Students were provided learning target logs, frameworks for portfolios, and reflection protocols. Teachers coached the students on setting learning goals, using self-assessment and articulating their progress to their peers and teachers. These data, generated by students, not only informed the students but also provided teachers with a wealth of information about what was happening in the classroom. In this module the parallel paths of teachers and students to becoming adaptive experts became explicit.

After completing the first two modules, teachers had much of the knowledge and many of the skills necessary to become a routine expert. Specifically teachers must have the comprehension and overall flexibility with their specific subject content as well as the subject specific pedagogy to build a culture of success and exploration in their STEM classrooms. In Module 3, teachers began to be elevated into adaptive experts who are able to solve new classroom challenges.

In Module 4, teachers developed their own research question and conducted an action research project that builds on the data collection and reflective practice of the first three modules (see for example, Gningue, Schroder & Peach, 2014). Research requires the formulation of questions based on observation. This is fundamental to the creative problem-solving inherent in becoming an adaptive expert. Transitioning to the role of researcher in their own classroom fostered more analytic teaching as well as a stronger commitment to the continuation their own professional development and collaboration with peers (Henson, 1996; Bransford, Brown, & Cocking, 1999). In these action research projects, teachers developed strategies that increase student engagement and learning in the STEM disciplines and then collected evidence of the effectiveness of these strategies. Developing research questions that provided new insight into classroom practice increased the flexibility that teachers brought to their own classroom; increased the professional standing of a teacher; added to the fund of teaching knowledge in the profession; and established teacher development and classroom practice as belonging to the classroom teacher (Bransford, Brown, & Cocking, 1999).

**MAST Rubrics**

Standardized rubrics were developed in the MAST system to guide coaches in the classroom as they observed, assessed and advised effective classroom practice. The rubric scores provided formative feedback for the teachers as they progressed through a module. On each module rubric there were approximately ten items (see Appendix A for examples). Accompanying each module rubric was a document that further articulated each item. Typically, each item was accompanied by a description, a longer note that described how it could be implemented in a classroom, and one to four teaching strategies a teacher could use to produce the desired student engagement and achievement. This detailed document for each item provided suggestions for teachers as they planned their classes and was used to ensure that the coaches provided reliable and consistent feedback to the teachers. Rubric items were scored on a scale of one to five according to the criteria in Appendix B. Rubric scores depended not only on the classroom behaviors of teachers and students, but also on the teachers' usage of data-driven reflection and metacognition outside of the classroom.

The coaches responsible for using the rubrics with teachers underwent extensive training. Prior to scoring MAST teachers, coaches scored multiple videos and sample classrooms using the rubric. The coaches' scores were compared to the rubric developers' and differences resolved. Often these discussions led to rubric revisions and clarifications. By the time that the MAST System was implemented, rubric use produced scores with high-levels of agreement. Periodically in the two-year period of the implementation described here, the rubric developers independently scored teachers to ensure that the coaches were continuing to adhere to the rubric.

Each teacher was evaluated once a month using student data and a classroom observation, which was evaluated by a rubric. Before the one-on-one session with the evaluating coach, a teacher evaluated their own performance using the same rubric. When the teacher and coach met, they compared rubric scores and used this as well as student data as a starting point for discussion. The rubric forms included spaces to record an individual action plan. The rubric scores also guide the coach in determining if a teacher was prepared to progress to the next module. Generally a rubric average of three or more indicated that a teacher was prepared.

Averages of all of the teacher scores at a school provided a measure of both a school's implementation of research-based strategies and teachers' usage of student data to make decisions. The schools with higher rubric averages had teachers who were progressing toward becoming adaptive teachers by employing metacognition. The schools with lower averages were populated with teachers who were not demonstrating the skills of an adaptive expert.

**MAST Cycle**

Constructivist theories of learning suggest that successful teacher professional development should utilize a learning cycle structure (Borasi & Fonzi, 2002). Following this constructivist approach, and the experiential learning cycles suggested by Kolb (1984), each MAST Module was built around monthly MAST Cycles. In a cyclic program, teachers begin with the knowledge they have and gradually, through multiple iterations of the MAST Cycle, expand their knowledge. In most modules there were four iterations of the MAST Cycle. The teachers focused on real-world problems that they faced in their classroom, rather than theoretical abstractions (Lave, 1988). Teachers continually applied what they were learning to their own classes and evaluated what worked for them and their students.

In the first phase of the cycle, called the Assimilation phase, teachers learned new classroom strategies via demonstrations, another key component of effective learning that complements the experiential learning (Merrill, 2002). The cycle progressed through a phase of reflective observation (Organizing Evidence), abstract conceptualization (Analyzing Evidence) and active experimentation (Processing & Planning) (Figure 1).  We did not show the concrete experience in the teachers' classrooms in the illustration as it occurred continuously since teachers continued to teach as they were engaged in the MAST System.

Teachers were guided toward becoming routine experts as they learned and applied specific new strategies in the classroom. To become adaptive experts, the teachers had to learn how to be self-reflective and critically analyze their own performance. To accomplish this goal, the metacognitive Defense of Instruction was added to all phases of the MAST Cycle in the form of a cognitive apprenticeship.

As with a traditional apprenticeship, the expert (here a coach) modeled their practice for the learner. As the learner (here a teacher) began to practice the necessary skills, the expert coached and provided scaffolding to support the learner. This modeling, coaching and scaffolding helped the teacher develop domain knowledge as well as the necessary self-monitoring and control strategies. Unlike a traditional apprenticeship, cognitive apprenticeships focus on mental tasks that are largely unseen (Collins, Brown, & Newman, 1989). To model the unseen, the coach had to articulate their thought process and decision-making. Similarly, for the coach to provide feedback or effective scaffolding, the teacher also had to articulate their thought process.

The process, in which teachers were asked to articulate their specific decisions and the reasons they made them, is the Defense of Instruction. The goal wasn't simply to defend past decisions, but rather to learn how to make the intentional decisions necessary to become adaptive experts. Teachers articulated their class goals and objectives and their progress toward them. They based their self-evaluations on objective student achievement and performance data. Coaches asked questions that challenged the teachers to refine their goals and explore alternative methods of accomplishing them. Sample questions that teachers addressed are shown in each phase of the MAST Cycle in Figure 1. The questions that coaches ask are the kinds that teachers will eventually incorporate into their own reflective practice.

It was in this process of Defense of Instruction that teachers learned how to approach their practice metacognitively. They monitored progress toward goals, monitored their own development, and received and analyzed feedback. In addition to being coached on their own decision-making, teachers were able to witness exemplary models of decision-making. The coaches articulated their thought process on similar decisions that they had made in their own classrooms. Defense of Instruction in the initial modules reinforced the skills teachers needed to engage in independent action research in Module 4. The focus on self-monitoring and reflection allowed teachers to engage metacognitively and make appropriate adjustments in their teaching. Below we describe each of the four phases of the MAST Cycle and how Defense of Instruction was integrated within each.

**Assimilating.** Seminars were held at the school site at the beginning of each monthly cycle. These seminars frequently began with a mathematics specific classroom problem. Solving this problem together served as a focal point for subsequent discussion. Becoming an adaptive expert in a STEM field requires superior subject specific knowledge that allows for flexible instructional approaches and the ability to apply meaningful context for students (Ball & Bass, 2000). "High knowledge teachers" can use their deep subject area knowledge to support mathematically or scientifically rigorous explanations and articulate sophisticated discipline-based reasoning (Hill, et al., 2008). Without this critical ingredient, opportunities for students to deepen their understanding of mathematics and science may be left unexplored and unprobed (Ball & Bass, 2000). Coaches also presented research-based classroom strategies and discussed how to effectively implement these strategies. Time was allocated in each seminar session for processing and teacher planning, so by the end of each seminar teachers had concrete lesson plans as well as a blueprint of the student data to be collected to use in the next session in customizing strategies. Table 1 lists MAST activities and provides an estimate of the duration of each.

During Instructional Rounds, participating teachers observed the strategies introduced in the seminar. Using the reflection/defense process, as teachers observed a strategy modeled, they collected evidence of its success and then collaboratively analyzed the practice within the context of their own classroom. This helped them make decisions about how to integrate the best practices into their own classes and to articulate this decision. (City, Elmore, Fiarman, & Teitel, 2009). Via Defense of Instruction, coaches asked teachers to reflect on which strategies would best help their students and why, and together they planned what evidence the teachers could collect to support that conjecture.

**Organizing Evidence.** Active Apprenticeship was provided at the participating teacher's site (Beard & Wilson, 2013). Using the MAST rubrics as a guide, coaches continued to help teachers execute a teaching strategy, which was introduced at the workshop and observed in the instructional rounds, in their classrooms as well as collect data on its implementation in the classroom. This allowed them to provide real-time coaching to improve implementation. Teachers used multiple data sources to evaluate student engagement and achievement. This triangulation of data allowed teachers to be informed when reflecting and making future decisions.

**Analyzing Evidence.** In Debriefing Sessions, coaches provided teachers with one on one metacognitive and blended coaching as described in Bloom, Castagna, Moir, and Warren (2005). These sessions included opportunities for teachers to practice metacognitive skills inherent in the Defense of Instruction, as well as analyze their progress within their assigned module. It allowed them to develop an action plan to strengthen their practice. The analysis and planning was done through the lens of student evidence, data was gathered both by the coach in the Active Apprenticeship phase and the teacher in the classroom. Teachers analyzed their own implementation prior to the debrief by examining student evidence that they collected.

**Processing and Planning.** Once a cycle, coaches conducted Co-planning Sessions in the field, or virtually, with a small (6-7 teachers) group. In these sessions, coaches and teachers collaborated to overcome obstacles that were preventing a shift in teaching and learning practices and also designed new resources that engaged all students in learning. This social aspect also followed from the cognitive apprenticeship model on which Defense of Instruction was built. By having coaches and teachers work together in groups, teachers were able to assess their own development in light of their peers' progress. Working together gave teachers an on-site community to support their classroom, provide readily accessible resources and cement their own professional development. This helped sustain and magnify the classroom innovation.

In addition to these group sessions teachers also worked individually with their coach to address specific concerns. These One-on-One Assistance with Preparing sessions were an opportunity for teachers to customize strategies to use in their classes, prepare for the next student learning targets, and plan future data collection. While the teachers were the ones making the decisions, the role of the coach was to ask probing questions and offer suggestions via the Defense of Instruction process. In all of the Modules, but especially the initial ones, the coach used this one-on-one opportunity to model their decision making by articulating how they would approach the teacher's concerns and use data to decide on a course of action.

**Other Activities.** In addition to the activities included within the MAST Cycle there were several other components, which occurred less frequently, but were also critical to the MAST System's success. Summer Workshops (held at the university) initially introduced the MAST System and then laid the foundations of collaboration by bringing together STEM teachers, coaches, and STEM faculty. The focus of these sessions was to share recent research, education trends, and policy recommendations. The teachers also worked collaboratively with the coaches and STEM faculty to establish student learning goals for the academic year.

University STEM faculty collaborated with coaches to present Quarterly STEM Learning Seminars. These sessions strengthened teachers' academic expertise as well as their specialized knowledge for STEM teaching. STEM faculty also provided insight into the expectations in undergraduate STEM courses. The focus of these seminars was a STEM problem that required deep understanding, such as a SAT math or AP Calculus question that was released by College Board. Participants analyzed the mathematics or scientific knowledge needed to effectively design appropriate learning progressions for their particular students and course level. These sessions shared many of the characteristics of other highly metacognitively-rich programs, see, for example the "meta-talk" Superfine and Li (2014) incorporate into their workshop format.

**MAST System Effectiveness**

The MAST system was first implemented in seven Alliance for College-Ready Public Schools (Alliance) in the summer of 2007 for academic year 2007–2008 (AY '07–08). Alliance College-Ready Public Schools is a nonprofit charter management organization operating schools in the Los Angeles Unified School District (LAUSD). In AY '06–07 Alliance managed six high schools and one middle school. Alliance added one new secondary school in AY '07–08 and five in AY '08–09, all eventually participated in the MAST System. As a partial description of the student populations of the two organizations, we note that at the start of the MAST intervention, 76.3% of the LAUSD students and 95.0% of the Alliance students qualified for free or reduced lunches (http://dq.cde.ca.gov/dataquest/). Here, we focus on the seven schools that were part of the system in AY '06–07, although we report some aggregated data that includes all Alliance schools. Forty-five teachers and over 11,000 students participated in the two-year intervention.

Assessing the effectiveness of professional development fundamentally rests on assessing teacher quality and there is little agreement on the best approaches to identifying and assessing teacher quality (Bolyard and Moyer-Packenham, 2008; Goldhaber and Anthony, 2007) so it is no surprise that there is no consensus on assessing professional development programs (see, for example, Darling-Hammond (2015); Briggs & Domingue (2011); or Bolyard & Moyer-Packenham, 2008). Researchers in the field often use teacher-related variables such as changes in understanding or attitude; however, there must be large changes of teacher-related variables in order to correlate with even small changes in student learning (Quint, 2011). For us, the obvious gold standard for quality education is student success, but there is also no universally accepted measure of student success (Moyer-Packenham et al., 2008, for review). We favor using a direct measure of student outcomes in the form of the high-stakes standardized tests that were in place in most schools due to the implementation of No Child Left Behind Act of 2001 (2002). Although inherently problematic in assessing individual teacher performance, these tests have the potential to be useful in evaluating program change. They are administered frequently and data can be collected retroactively. Thus, they are positioned to measure effects quickly, compared to other important measures like teacher retention, student retention, and student success in STEM disciplines. Quint (2011) warns, however, that it may be unrealistic to expect to see major shifts in student achievement in a year or two. On the other hand, observed effects are likely to be important.

We used normalized changes in student performance on the California Standards Test (CST) in mathematics to assess the effect of the MAST System at each school site. Using normalized change <$c$> minimizes the effect of variations in initial scores on changes in performance (Marx & Cummings, 2007). The CST was given in May to California public school students, grades 2–11, including charter schools like Alliance. High school students took the CST in general math, algebra I, geometry, algebra II, or summative high school math, depending on their math enrollment. The CA Board of Education reported CST scores as a percentage of students in each of the following five categories: Advanced, Proficient, Basic, Below Basic or Far Below Basic, where Basic is the minimum acceptable performance. A positive normalized change indicates that the number of students in a particular category has increased. This would be a desirable outcome in the number of students in the Advanced or Proficient categories. A negative normalized change indicates that the number of students in a particular category has decreased. This would be a desirable outcome in the lower categories, Below Basic or Far Below Basic.

**Impact of the MAST System**

All seven Alliance schools in the study had CST scores for AY '06–07 (CST '07) before the implementation of the MAST System, as well as scores midway through the two-year program (CST '08) and at the end (CST '09). After the implementation of MAST, these seven Alliance schools increased the percentage of students who were well-prepared in mathematics and most reduced the percentage of students who were inadequately prepared in mathematics (Figure 2). Some schools showed a remarkable improvement in only two years. For example, School 4 (Figure 2) increased the number of students who scored at or above Proficient from 14% to 53% ($<c>= 0.45$) and reduced those below Basic from 58% to 25% ($<c>= –0.57$) (Table 2).

As is common in educational interventions, the MAST system was not implemented as part of an experimental protocol designed to test the effect of the system. It was implemented in response to the Alliance schools desire to improve their STEM education. Several lines of evidence suggest, however, that these improvements were related to the implementation of the MAST System. These include: the absence of a similar improvement in the mathematics CST in the district or the state; the absence of similar improvement in the non-STEM CST scores in the seven Alliance schools; and the relationship between the degree to which the teachers implement adaptive expert strategies and the size of the normalized gains of the students on the CST.

First, the upward trend in scores was not seen in either the district or the state. In Figure 3 we report the mathematics scores from all California schools, all LAUSD schools and the seven Alliance schools. The dramatic rise in the number of students at or above Proficient in the Alliance schools (11% to 29%, $<c>= 0.20$) was not reflected in the data from the state (31% to 33%, $<c>= 0.03$) or the district (19% to 24%, $<c>= 0.06$) (Figure 3). In 2007, a larger percentage of Alliance students (68%) were below Basic than in either the state (41%) or district (58%). At the seven Alliance schools there was a dramatic drop in the number of these inadequately prepared students to 46% in 2009. This meant that the schools in which the MAST System was implemented had a lower percentage of students below Basic than the state (53%) or district (52%) and a better normalized change than either (–0.32 versus –0.10 for LAUSD and +0.20 for CA) at the end of the two-year intervention.

Second, like Hill, Rowan and Ball (2005), we used the scores on the English Language Arts (ELA) exam as an internal marker. The impressive normalized changes in mathematics scores were not paralleled by improvements in the ELA CST scores at the Alliance schools. Student success can be influenced by a variety of external factors including, for example, the changes in student population that might be expected at relatively new urban schools. If external factors unrelated to the MAST System accounted for improved scores, similar improvements should have manifested in CST scores in non-STEM disciplines. What is observed over this two-year period is that the improvements on the CST ELA test among the seven Alliance schools was similar to those of the state and district (Figure 4). The seven Alliance schools improved the number of students scoring above Basic ($<c> = 0.08$), as did the district ($<c> = 0.07$) and state ($<c> = 0.12$). All three organizations experienced similar decreases in the number of students scoring below Basic (Alliance $<c> = –0.19$, district $<c> = –0.16$ and state $<c> = –0.17$). The fact that the seven Alliance schools did not have larger improvements in the CST ELA scores than the other organizations' suggested the Alliance student populations were not changing more rapidly than the others, and the major influence on the Alliance mathematics CST scores was the intervention with the MAST System.

Third, there was a relationship between the normalized gain in student performance and the degree to which teachers implemented research-based strategies and used of data-driven reflection, as indicated by scores on the MAST Rubrics (Appendix A). The MAST rubric scores, integral to the MAST learning cycle, generally went up as a teacher progresses through the system indicating that rubric scores were not just a measure of pre-existing teacher quality. The schools in Figure 2 are arranged from left to right on the basis of the average MAST rubric scores (from high to low), which describe the degree to which teachers were behaving like adaptive experts. The average rubric score for the two schools on the left of the graph (roughly representing the top quartile) was 2.3 and the two schools on the right (bottom quartile) was 0.8. Schools 1 and 2 had large average normalized changes (0.18 for students at or above Proficient and –0.34 for those below Basic); Schools 6 and 7 showed almost no change in either category (normalized changes of 0.03 and –0.01 respectively). The schools where teachers had low levels of implementing adaptive expert strategies showed little or no change in student performance.

**Sustained Improvements**

Central to the MAST System was the premise that if the teachers can acquire the skills of an adaptive expert, they will be able to continue improving after the two-year intervention. To examine this, we further studied the student CST scores from the five schools (#1-5) that showed improvement while participating in the MAST System. For these schools, the mathematics CST scores were compiled for two years following their MAST System intervention. Figure 5 shows the scores over a five-year period that spans the year prior to their involvement, the two-year MAST System participation and the two years immediately after their participation. It was apparent from Figure 5 that all of the schools that demonstrated significant implementation of adaptive expert strategies were able to retain the student achievement gains that they had made.

More importantly, four of the schools continued to improve their student achievement scores. The number of students scoring Advanced or Proficient continued to increase after their involvement in MAST. Likewise, the number of students scoring Below Basic or Far Below Basic continued to decrease. Across all five schools, the average normalized changes in these two student populations from the final year of MAST participation to two years later were +0.15 (Advanced or Proficient) and –0.33 (below Basic). While not as large as the normalized changes that were experienced during the two-year MAST System intervention (Table 2), these were improvements on top of those already seen as a result of the intervention. They were positive trends that indicate that teachers were able to further improve their practice. It should be noted that the reported CST scores were for all students and teachers at a school, regardless of their MAST participation. Given the inevitable turnover among teachers, it would be reasonable to see less than optimal student scores as new teachers, without MAST training, were hired after the school has completed the two-year MAST system intervention. Based on these data, we believe that the MAST system had a significant effect not only on student performance but also the individual teachers and/or the culture of the school such that the improvements experienced during the intervention not only persisted but continue to improve after the intervention.

## Discussion

The Mathematics and Science Teaching (MAST) System resulted in desirable changes in the CST scores of students at most schools where it was implemented. This professional development system had a number of important components, which are generally shared by successful professional development programs. The MAST System used workshops with outside experts to provide both pedagogy and subject matter expertise (Guskey & Yoon, 2009). The curriculum was intensive with more than eighty hours in each year of the intervention (Table 1) (Guskey & Yoon, 2009). It combined a variety of approaches both on site and off site that provided teachers with coaching and scaffolding for an extended duration. Throughout, it used an experiential learning cycle to reinforce and integrate teaching strategies into teachers' classroom practices.

Although all of these features were important to its success, it is our opinion that the success of the MAST System is primarily due to the metacognitive practices interwoven into every aspect of the program. In every cycle of every module teachers were asked to actively engage and reflect meaningfully. Much of this reflection occurred in the Defense of Instruction where teachers defended each decision they made for their classrooms with real data collected from their own experience. While this defense process was initially unfamiliar to many teachers, the MAST coaches modeled this process and provided personalized coaching and scaffolding to each participant.

This continuous review and reflection established a habit of metacognition that carried forward so that improvements in the classroom were not only sustained, but continued to grow. Teachers were able to respond to situations with innovative strategies and knew how to test them for effectiveness. In this way, the teachers became adaptive experts who were well-prepared for any new situation that they may encounter in the classroom. We believe that it is this development of adaptive experts through metacognitive practices that was responsible for the success of the MAST System.

METACOGNITIVE TRAINING IN PROFESSIONAL DEVELOPMENT    27*School District teachers by the Los Angeles Times*. Boulder, CO: National Education Policy Center.

City, E. A., Elmore, R. F., Fiarman, S. E., & Teitel, L. (2009). *Instructional rounds in education: A network approach to improving teaching and learning*. Cambridge, MA: Harvard University Press.

Collins, A., Brown, J.S., & Newman, S.E. (1989). Cognitive apprenticeship: Teaching the crafts of reading, writing, and mathematics. In L. B. Resnick (Ed.) *Knowing, learning, and instruction: Essays in honor of Robert Glaser* (pp. 453-494). Hillsdale, NJ: Lawrence Erlbaum Associates.

Committee on STEM Education. (2013). *Report to the President: Federal Science, Technology, Engineering, and Mathematics (STEM) Education 5-Year Strategic Plan*. Retrieved from https://www.whitehouse.gov/sites/default/files/microsites/ostp/stem_stratplan_2013.pdf

Darling-Hammond, L. (2015). Can Value Added Add Value to Teacher Evaluation? *Educational Researcher*, 44 (2), 132–137.

Darling-Hammond, L., Holtzman, D., Gatlin, S., & Vasquez Heilig, J. (2005). Does Teacher Preparation Matter? Evidence about Teacher Certification, Teach for America, and Teacher Effectiveness. *Education Policy Analysis Archives*, 13 (42), 1-48.

Darling-Hammond, L., Wei, R. C., Andree, A., Richardson, N., & Orphanos, S. (2009). State of the Profession: Study Measures Status of Professional Development. *Journal Of Staff Development*, 30(2), 42-44.

DeSilver, D. (2015). *U.S. students improving – slowly – in math and science, but still lagging internationally*. Retrieved from http://www.pewresearch.org/fact-tank/2015/02/02/u-s-students-improving-slowly-in-math-and-science-but-still-lagging-internationally/

METACOGNITIVE TRAINING IN PROFESSIONAL DEVELOPMENT                                28Farmer, J., Gerretson, H., & Lassak, M. (2003). What teachers take from professional development: cases and implications. *Journal of Mathematics Teacher Education*, 6, 331–360.

Flavell, J. H. (1979). Metacognition and cognitive monitoring: A new area of cognitive–developmental inquiry. *American psychologist,* 34(10), 906-911.

Funk, C., Rainie, L & Page, D. (2015). *Public and Scientists' Views on Science and Society*. Retrieved from http://www.pewinternet.org/files/2015/01/PI_ScienceandSociety_Report_012915.pdf

Gningue, S. M., Schroder, B., & Peach, R. (2014). Reshaping the "glass slipper": The development of reflective practice by mathematics teachers through action research. *American Secondary Education*, 42(3), 18-29.

Goldhaber, D., & Anthony, E. (2007). Can teacher quality be effectively assessed? National board certification as a signal of effective teaching. *The Review of Economics and Statistics*, 89(1), 134-150.

Goldhaber D. D. & Brewer, D. J. (1997). Why don't schools and teachers seem to matter? Assessing the impact of unobservables on educational productivity. *The Journal of Human Resources*, 32, 505–523.

Guskey, T. R., & Yoon, K. W. (2009). What Works in Professional Development? *Phi Delta Kappan*, 90 (7), 495-499.

Hammerness, K., Darling-Hammond, L., Bransford, J. (2005). How teachers learn and develop. In L. Darling-Hammond & J. Bransford (Eds.), *Preparing teachers for a changing world: what teachers should learn and be able to do* (pp. 258-289). San Francisco, CA: Jossey-Bass.

METACOGNITIVE TRAINING IN PROFESSIONAL DEVELOPMENT                31Mullis, I. V. S., Martin, M. O., Foy, P., & Arora, A. (2012). *TIMSS 2011 International Results in Mathematics*. Chestnut Hill, MA: Boston College.

The Nation's Report Card. (2015). *U.S. Department of Education, Institute of Education Sciences, National Center for Education Statistics*. Retrieved from http://www.nationsreportcard.gov/

National Academy of Sciences, National Academy of Engineering and Institute of Medicine. (2007). *Rising Above the Gathering Storm: Energizing and Employing America for a Brighter Economic Future*. Washington, DC: The National Academies Press.

National Academy of Sciences, National Academy of Engineering, and Institute of Medicine. (2010). Rising Above the Gathering Storm, Revisited: Rapidly Approaching Category 5. Washington, DC: The National Academies Press.

No Child Left Behind Act of 2001. (2002). Pub. L. No. 107-110, § 115. *Stat, 1425*, 107-110.

Olson, S., & Riordan, D. G. (2012). *Engage to Excel: Producing One Million Additional College Graduates with Degrees in Science, Technology, Engineering, and Mathematics. Report to the President*. Executive Office of the President.

Organisation for Economic Co-operation and Development. (2014). *PISA 2012 Results: What Students Know and Can Do (Volume I, Revised Edition, February 2014): Student Performance in Mathematics, Reading and Science*. OECD Publishing.

Price, M. (2013). *A Downstream Pathway into Teaching*. Retrieved from http://www.sciencemag.org/careers/2013/04/downstream-pathway-teaching

Quint, J. (2011). *Professional development for teachers: What two rigorous studies tell us*. NewYork, NY: Manpower Demonstration Research Corporation. Retrieved from http://www.mdrc.org/publication/professional-development-teachers

Appendix A

Sample Items from Module Rubrics

Module 1: MAST Culture of Trust and Success

- Students engage in inquiry through an activity that appropriately connects past and present academic content within the frame of a story.
- Students engage in the lesson as the teacher connects it to previous content and illuminates the part of the story being told by the teacher around essential new content.
- Students engage in content appropriate higher cognitive questions that involve individual thought and collaboration with adequate wait time on the essential content.

Module 2: MAST Culture of Collaboration and Persistence

- Students have an opportunity to gather and use evidence to demonstrate achievement of the unit learning targets.
- Students view mistakes as a part of learning and use reflective error analysis to think critically about their work measured against the expectations for mastery.
- Students embrace an environment where calculated risk-taking is a valued asset in assessment for learning.

Module 3: MAST Culture of Conditionalized, Flexible and Fluent Knowledge

- Students and the teacher actively connect their understanding/interpretation of mathematical or scientific concepts to deepen student understanding through conditionalizing knowledge by making conjectures and justifying conclusions.
- Students become adaptive experts in mathematical or scientific knowledge and use assessments to advocate for their learning needs to develop flexibility and fluency of mathematical and scientific knowledge.
- Students engage in a learning culture that cultivates adaptive experts who view mathematics and scientific knowledge as conditional, flexible and fluent.

Module 4: MAST Culture of Action Research

In Module 4, teachers' progress was evaluated by the produce of their action research project, rather than a rubric applied to their classroom. Teachers were required to make a public presentation of their action research project.

Appendix B

MAST Rubric Criteria

5. *Expert Implementation*– A MAST practice is effectively implemented when:
- 80% of the students are actively engaged and/or demonstrating achievement;
- the teacher is actively using classroom data to design/adapt lessons;
- a 1-to-1correspondence exists between auditory and visual communication;
- the practice is accurately implemented per MAST guidelines;
- academic content is appropriate, rigorous, chunked, and embeds re-teaching.

4. *Proficient Implementation*– A MAST practice is proficiently implemented when:
- 70% of the students are actively engaged and/or demonstrating achievement;
- the teacher is actively using classroom data to design/adapt lessons;
- a 1-to-1 correspondence <u>does not exist</u> between auditory and visual communication;
- the practice is accurately implemented per MAST guidelines;
- academic content is appropriate, rigorous, chunked, and embeds re-teaching.

3. *Emerging Implementation*– A MAST practice is adequately implemented when:
- 60% of the students are actively engaged and/or demonstrating achievement;
- the teacher is <u>not</u> actively using classroom data to design/adapt lessons;
- a 1-to-1 correspondence <u>does not exist</u> between auditory and visual communication;
- the practice is accurately implemented per MAST guidelines;
- academic content is appropriate, rigorous, chunked, and embeds re-teaching.

2. *Engaging Implementation*– A MAST practice is minimally implemented when:
- 50% of the students are actively engaged and/or demonstrating achievement;
- the teacher is <u>not</u> actively using classroom data to design/adapt lessons;
- a 1-to-1 correspondence <u>does not exist</u> between auditory and visual communication;
- the practice is <u>not accurately implemented</u> per MAST guidelines;
- academic content is appropriate, but is not rigorous or chunked, and <u>does not</u> embed re-teaching.

1. *Beginning Implementation*– A MAST practice is ineffectively implemented when:
- less than 50% of the students are actively engaged and/or demonstrating achievement;
- the teacher is <u>not</u> actively using classroom data to design/adapt lessons;
- a 1-to-1 correspondence <u>does not exist</u> between auditory and visual communication;
- the practice is <u>not accurately implemented</u> per MAST guidelines;
- academic content is <u>not</u> appropriate, rigorous, and/or chunked, and <u>does not</u> embed re-teaching

0. *Missed Opportunity–* A missed opportunity for a MAST practice occurs when:
   - Use of a MAST practice within the flow of the lesson would have increased student engagement
   - Use of a MAST practice within the flow of the lesson would have increased student understanding of the content

N/A. *Not Applicable–* A MAST practice is not applicable when:
   - Use of a MAST practice is not a focus for the month
   - Use of a MAST practice is not appropriate within the flow of the lesson

Table 1

Sample Year of MAST Activities and Durations

| Activity | Number of Days | Number of Hours |
|---|---|---|
| Summer Workshops introduced teachers to the suggested MAST practices and facilitated the collaborative development of learning targets and sequencing. | 2 | 12 |
| Quarterly Mathematics Learning Seminars brought together coaches, teachers and STEM faculty with the goal of strengthening teachers' academic expertise. | 3 | 6 |
| **Assimilation** Seminars at schools introduced teachers to new classroom research-supported strategies. With education and STEM faculty, teachers engaged in a STEM topic in order to effectively design and teach that topic | 12 | 24 |
| Instructional Rounds allowed teachers to observe and analyze classrooms where MAST strategies were successfully implemented. | 4 | 12 |
| **Organizing Evidence** Active Apprenticeship occurred at the school site via classroom observations and real-time coaching. Focus was on strategies that were previously introduced. | 10 | 10 |
| **Analyzing Evidence** One-on-one De-briefing Sessions where a teacher and coach examined what happened in the classroom. | 10 | 10 |
| **Processing and Planning** Co-planning Sessions were where coaches worked with small groups of teachers to solve problems in the classroom. | 10 | 10 |
| One-on-One Assistance with Preparing was where a coach member assisted a teacher in determining the next steps to improve their practice and plan a future lesson. | 10 | 10 |
| Total | 61 | 94 |

Table 2

Normalized Change of Participating Schools' CST Scores Over the Two-year MAST Intervention

| CST categories | Participating Schools | | | | | | |
| --- | --- | --- | --- | --- | --- | --- | --- |
| | 1 | 2 | 3 | 4 | 5 | 6 | 7 |
| Proficient and Advanced | 0.17 | 0.18 | 0.21 | 0.45 | 0.27 | 0.03 | 0.03 |
| Below and Far Below Basic | –0.43 | –0.26 | –0.36 | –0.57 | –0.47 | 0.02 | –0.05 |

Note: To compare the improvement in the seven schools, which had different initial CST scores in 2007, we computed the normalized change, which ranges from –1 (maximum decrease) to +1 (maximum increase). For increases, the normalized change is computed by dividing the actual gain by the possible gain. For decreases, the actual loss is divided by the initial score. Normalized changes were computed for both the percent of students at or above Proficient as well as the percent below Basic at each school.

Figure 1

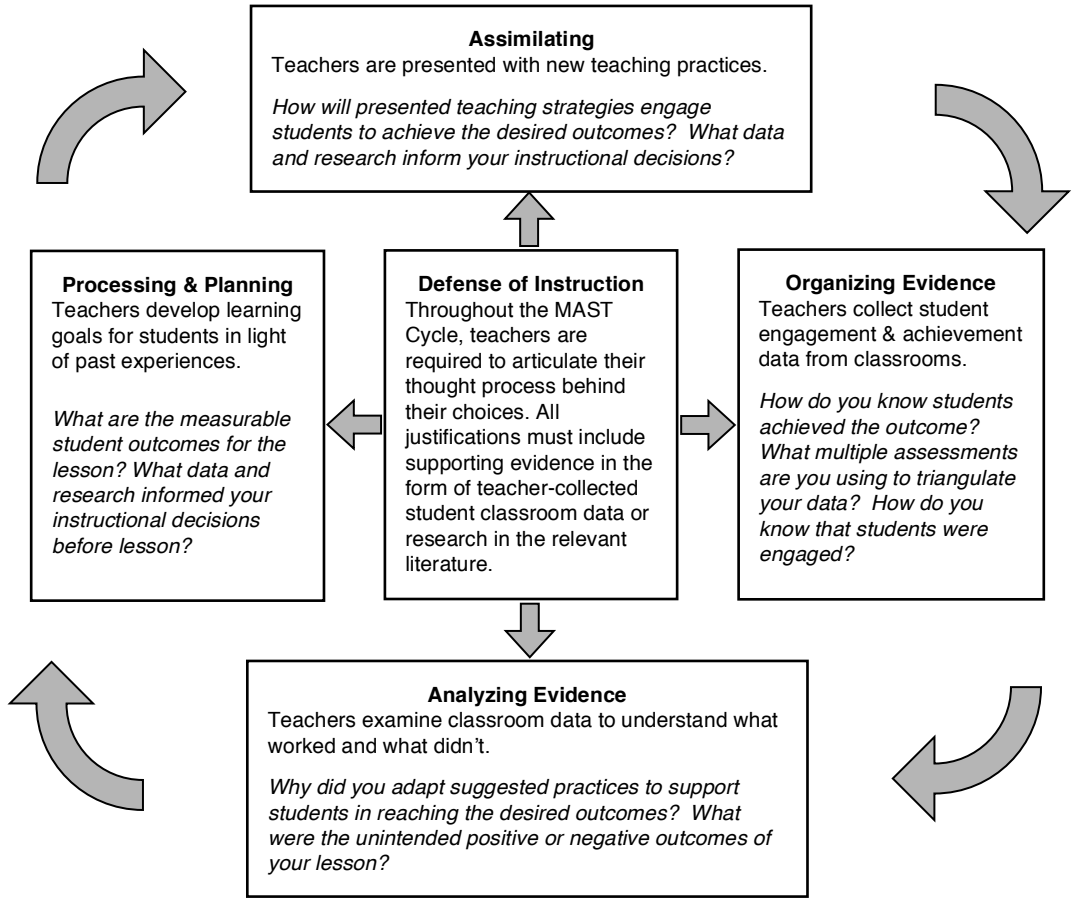

*Figure 1:* Schematic of the MAST Cycle. The MAST Cycle emphasized having teachers practice and receive feedback on the skills they needed to become self-regulated or adaptive teachers. This cycle repeats each month with a focus on different teacher practices and/ or student difficulties with STEM content. Central to each phase of the cycle were opportunities for teachers to develop these skills in the Defense of Instruction process. Sample questions of that process are shown in each phase.

Figure 2

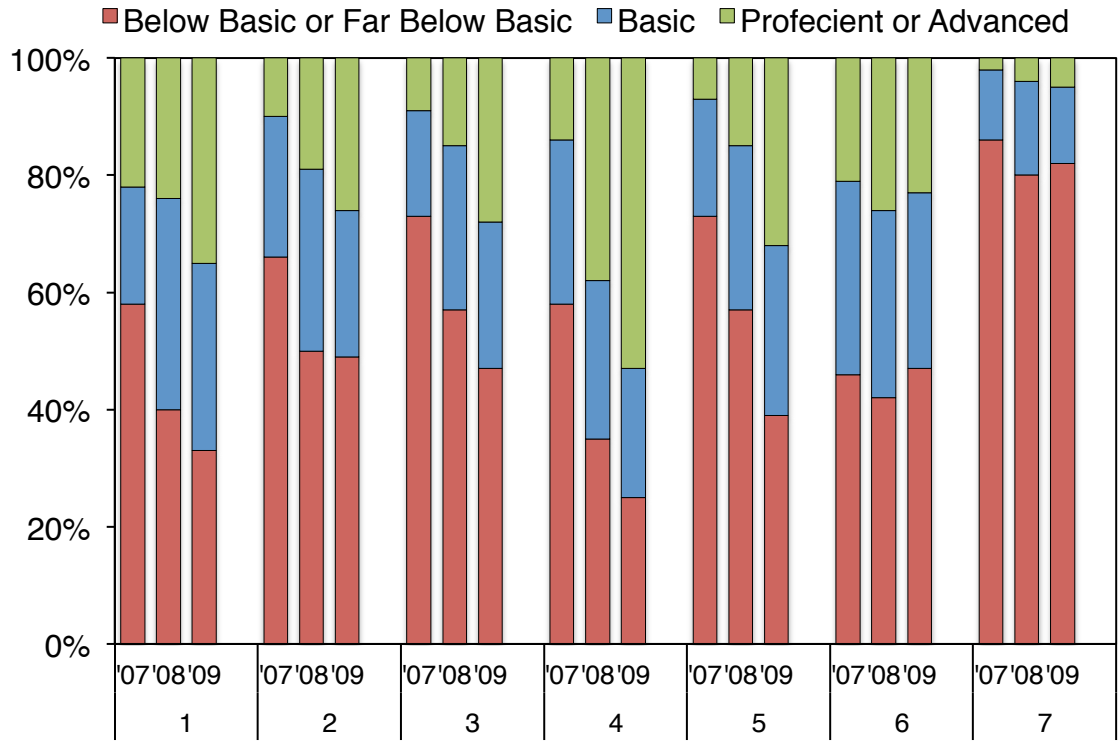

*Figure 2:* 2007–2009 CST Math scores for the seven Alliance secondary schools that participated in the MAST System. Data for each school site (1–7) are shown as percentage of students with scores below the state minimum (red), percentage of students scoring at the minimum level (blue) and percentage of students above the minimum (green). Schools are arranged by average rubric score decreasing from left to right (2.5, 2.1, 1.6, 1.4, 1.3, 1.0 and 0.6), rubric scores range from 0 (no evidence of adaptive expertise behaviors) to 5, (clear evidence of research-based strategies in the classroom and reflection based on student data).

Figure 3

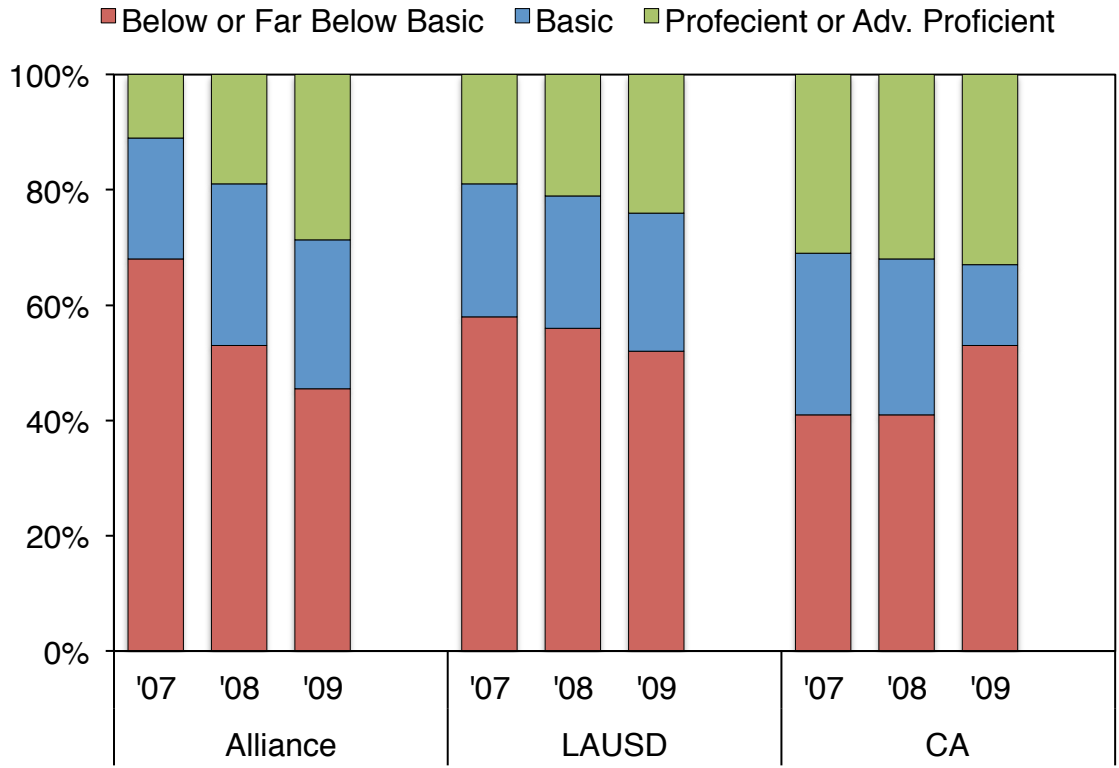

*Figure 3*: 2007– 2009 CST mathematics scores for the seven Alliance schools studied, LA Unified School District (LAUSD) and California (CA). From the year before the MAST intervention (2007) to the final year of the intervention (2009), there was a dramatic rise in the number of Alliance students at or above Proficient (11% to 29%, $<c>= 0.20$), which was not seen in the district (19% to 24%, $<c>= 0.06$) or state (31% to 33%, $<c>= 0.03$) populations. Similarly, there was a larger drop in the number of Alliance students below the Basic level (68% to 46%, $<c>=–0.32$), that what was seen in LAUSD (58% to 52%, $<c>=–0.10$). Over this time period, the state actually had an increase (41% to 53%, $<c>=0.20$) in the number of students scoring below Basic.

Figure 4

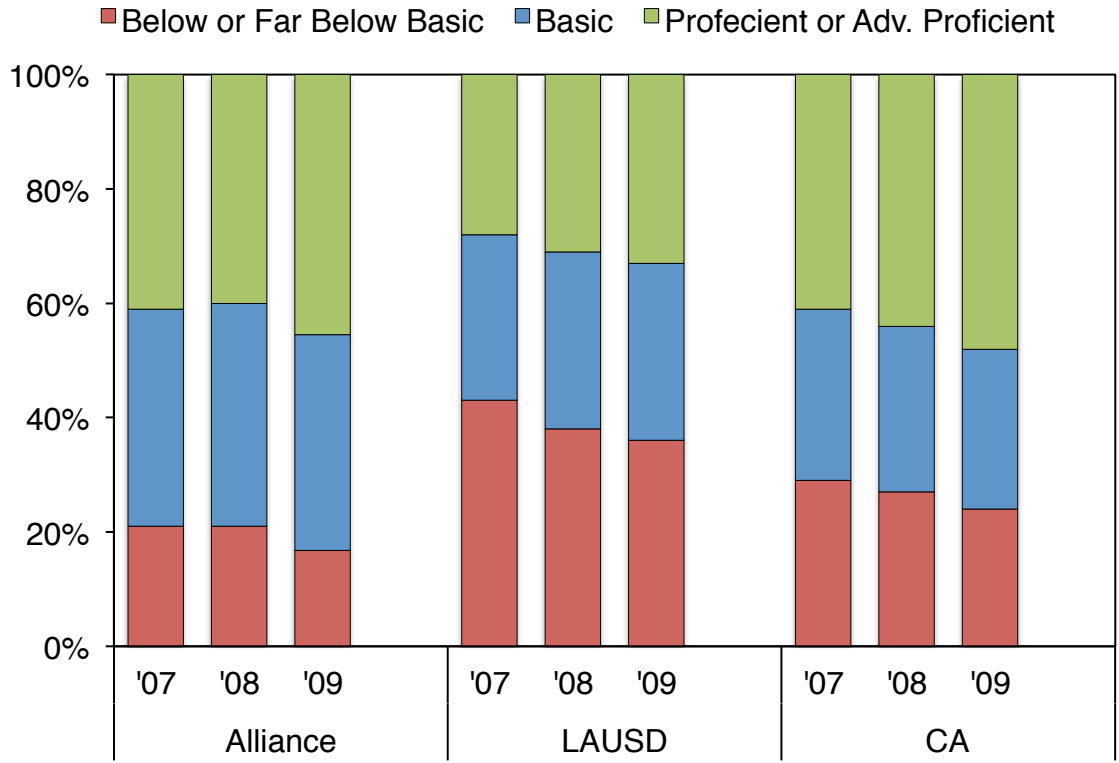

*Figure 4*: 2007– 2009 CST English language arts averages for the seven Alliance schools studied, LA Unified School District (LAUSD) and California (CA). From prior to the MAST intervention (2007) to its conclusion (2009), the number of students within the MAST trained schools who scored at or above Proficient increased from 41% to 46% ($<c>= 0.08$) and the number who scored below Basic decreased from 21% to 17% ($<c>= –0.19$). The district and state had comparable improvements in the number of students at or above Proficient (28% to 33%, $<c>= 0.07$, and 41% to 48%, $<c>= 0.12$, respectively) and the number below Basic (43% to 36%, $<c>=–0.16$, and 29% to 24%, $<c>=–0.17$, respectively). These similar changes on non-STEM tests suggest that the Alliance student population did not change with respect to the district or state populations and that the changes in CST mathematics scores were due to changes in the mathematics classrooms.

METACOGNITIVE TRAINING IN PROFESSIONAL DEVELOPMENT                                              43